\documentclass[12pt,a4paper,oneside]{article}
\usepackage[english]{babel}
\usepackage{amssymb}
\usepackage{latexsym}
\usepackage{graphicx}
\usepackage{subfigure}
\usepackage[dvips]{epsfig}

\begin{document}
\newcommand{\cc}[0]{${\rm c}\bar {\rm c}$}
\newcommand{\gtoc}[0]{${\rm g\rightarrow c \bar{c}}$}
\newcommand{\gtob}[0]{${\rm g\rightarrow b \bar{b}}$}
\newcommand{\zqq}[0]{${\rm Z\rightarrow q \bar{q}}$}
\newcommand{\zcc}[0]{${\rm Z\rightarrow c \bar{c}}$}
\newcommand{\zbb}[0]{${\rm Z\rightarrow b \bar{b}}$}
\newcommand{\zuds}[0]{${\rm Z\rightarrow u \bar{u},d \bar{d},s \bar{s}}$}
\def\qqbar{\mbox{${\rm q\bar{q}}$}}
\def\bbbar{\mbox{${\rm b\bar{b}}$}}
\def\ccbar{\mbox{${\rm c\bar{c}}$}}
\def\uubar{\mbox{${\rm{u\bar u}}$}}
\def\ddbar{\mbox{${\rm{d\bar d}}$}}
\def\ssbar{\mbox{${\rm{s\bar s}}$}}
\def\Z{\mbox{$\hbox{Z}$}}

\newcommand{\xEb}{\ensuremath{x_E^{\rm (b)}}}
\newcommand{\bl}{\ensuremath{{\rm b}\to \ell}}
\newcommand{\cl}{\ensuremath{{\rm c}\to \ell}}
\newcommand{\bcl}{\ensuremath{{\rm b}\to {\rm c}\to \ell}}
\newcommand{\xEc}{\ensuremath{x_E^{\rm (c)}}}
\newcommand{\gcc}{\ensuremath{g_{\rm c\bar c}}}
\newcommand{\gbb}{\ensuremath{g_{\rm b\bar b}}}
\newcommand{\Ne}{\ensuremath{N_{\rm e}}}
\newcommand{\Nmu}{\ensuremath{N_\mu}}

\begin{titlepage}
\pagestyle{empty}

\begin{center}
\rm EUROPEAN ORGANIZATION FOR NUCLEAR RESEARCH (CERN)
\end{center}

\vskip 1.0 cm

\begin{flushright}
{CERN-EP/2003-003} \\
January 30, 2003
\end{flushright}

\vskip 2.0 cm

\begin{center}
\boldmath
{\LARGE\bf A Measurement of the Gluon Splitting Rate into \cc\ Pairs in Hadronic Z Decays}
\vskip 1.0 cm
{\LARGE The ALEPH Collaboration $^{*)}$}
\end{center}
\vskip 2 cm
\centerline{\large\bf Abstract}
\vskip 0.6 cm
The rate of gluon splitting into \cc\ pairs in hadronic Z decays is measured using the data sample collected by ALEPH from 1991 to 1995.
The selection is based on the identification of leptons (electrons and muons) originating from semileptonic charm decays, and on the topological properties of signal events. The result derived from the selected sample is $\gcc =(3.26 \pm 0.23 \rm{(stat)} \pm 0.42  \rm{(syst)}) \%$.

\vskip 3.0 cm
\begin{center}
\it{Submitted to Physics Letters B}
\end{center}

*) See next pages for the list of authors.

\end{titlepage}
 
\setcounter{footnote}{0}
\setcounter{page}{1}
\topmargin=-1.cm

\pagestyle{empty}
\newpage
\small
%
%
\newlength{\saveparskip}
\newlength{\savetextheight}
\newlength{\savetopmargin}
\newlength{\savetextwidth}
\newlength{\saveoddsidemargin}
\newlength{\savetopsep}
\setlength{\saveparskip}{\parskip}
\setlength{\savetextheight}{\textheight}
\setlength{\savetopmargin}{\topmargin}
\setlength{\savetextwidth}{\textwidth}
\setlength{\saveoddsidemargin}{\oddsidemargin}
\setlength{\savetopsep}{\topsep}
%
%
\setlength{\parskip}{0.0cm}
\setlength{\textheight}{25.0cm}
\setlength{\topmargin}{-1.5cm}
\setlength{\textwidth}{16 cm}
\setlength{\oddsidemargin}{-0.0cm}
\setlength{\topsep}{1mm}
\pretolerance=10000
\centerline{\large\bf The ALEPH Collaboration}
\footnotesize
\vspace{0.5cm}
{\raggedbottom
\begin{sloppypar}
\samepage\noindent
A.~Heister,
S.~Schael
\nopagebreak
\begin{center}
\parbox{15.5cm}{\sl\samepage
Physikalisches Institut das RWTH-Aachen, D-52056 Aachen, Germany}
\end{center}\end{sloppypar}
\vspace{2mm}
\begin{sloppypar}
\noindent
R.~Barate,
R.~Bruneli\`ere,
I.~De~Bonis,
D.~Decamp,
C.~Goy,
S.~Jezequel,
J.-P.~Lees,
F.~Martin,
E.~Merle,
\mbox{M.-N.~Minard},
B.~Pietrzyk,
B.~Trocm\'e
\nopagebreak
\begin{center}
\parbox{15.5cm}{\sl\samepage
Laboratoire de Physique des Particules (LAPP), IN$^{2}$P$^{3}$-CNRS,
F-74019 Annecy-le-Vieux Cedex, France}
\end{center}\end{sloppypar}
\vspace{2mm}
\begin{sloppypar}
\noindent
S.~Bravo,
M.P.~Casado,
M.~Chmeissani,
J.M.~Crespo,
E.~Fernandez,
M.~Fernandez-Bosman,
Ll.~Garrido,$^{15}$
M.~Martinez,
A.~Pacheco,
H.~Ruiz
\nopagebreak
\begin{center}
\parbox{15.5cm}{\sl\samepage
Institut de F\'{i}sica d'Altes Energies, Universitat Aut\`{o}noma
de Barcelona, E-08193 Bellaterra (Barcelona), Spain$^{7}$}
\end{center}\end{sloppypar}
\vspace{2mm}
\begin{sloppypar}
\noindent
A.~Colaleo,
D.~Creanza,
N.~De~Filippis,
M.~de~Palma,
G.~Iaselli,
G.~Maggi,
M.~Maggi,
S.~Nuzzo,
A.~Ranieri,
G.~Raso,$^{24}$
F.~Ruggieri,
G.~Selvaggi,
L.~Silvestris,
P.~Tempesta,
A.~Tricomi,$^{3}$
G.~Zito
\nopagebreak
\begin{center}
\parbox{15.5cm}{\sl\samepage
Dipartimento di Fisica, INFN Sezione di Bari, I-70126 Bari, Italy}
\end{center}\end{sloppypar}
\vspace{2mm}
\begin{sloppypar}
\noindent
X.~Huang,
J.~Lin,
Q. Ouyang,
T.~Wang,
Y.~Xie,
R.~Xu,
S.~Xue,
J.~Zhang,
L.~Zhang,
W.~Zhao
\nopagebreak
\begin{center}
\parbox{15.5cm}{\sl\samepage
Institute of High Energy Physics, Academia Sinica, Beijing, The People's
Republic of China$^{8}$}
\end{center}\end{sloppypar}
\vspace{2mm}
\begin{sloppypar}
\noindent
D.~Abbaneo,
T.~Barklow,$^{26}$
O.~Buchm\"uller,$^{26}$
M.~Cattaneo,
F.~Cerutti,
B.~Clerbaux,$^{23}$
H.~Drevermann,
R.W.~Forty,
M.~Frank,
F.~Gianotti,
J.B.~Hansen,
J.~Harvey,
D.E.~Hutchcroft,$^{30}$,
P.~Janot,
B.~Jost,
M.~Kado,$^{2}$
P.~Mato,
A.~Moutoussi,
F.~Ranjard,
L.~Rolandi,
D.~Schlatter,
G.~Sguazzoni,
W.~Tejessy,
F.~Teubert,
A.~Valassi,
I.~Videau
\nopagebreak
\begin{center}
\parbox{15.5cm}{\sl\samepage
European Laboratory for Particle Physics (CERN), CH-1211 Geneva 23,
Switzerland}
\end{center}\end{sloppypar}
\vspace{2mm}
\begin{sloppypar}
\noindent
F.~Badaud,
S.~Dessagne,
A.~Falvard,$^{20}$
D.~Fayolle,
P.~Gay,
J.~Jousset,
B.~Michel,
S.~Monteil,
D.~Pallin,
J.M.~Pascolo,
P.~Perret
\nopagebreak
\begin{center}
\parbox{15.5cm}{\sl\samepage
Laboratoire de Physique Corpusculaire, Universit\'e Blaise Pascal,
IN$^{2}$P$^{3}$-CNRS, Clermont-Ferrand, F-63177 Aubi\`{e}re, France}
\end{center}\end{sloppypar}
\vspace{2mm}
\begin{sloppypar}
\noindent
J.D.~Hansen,
J.R.~Hansen,
P.H.~Hansen,
A.~Kraan,
B.S.~Nilsson
\nopagebreak
\begin{center}
\parbox{15.5cm}{\sl\samepage
Niels Bohr Institute, 2100 Copenhagen, DK-Denmark$^{9}$}
\end{center}\end{sloppypar}
\vspace{2mm}
\begin{sloppypar}
\noindent
A.~Kyriakis,
C.~Markou,
E.~Simopoulou,
A.~Vayaki,
K.~Zachariadou
\nopagebreak
\begin{center}
\parbox{15.5cm}{\sl\samepage
Nuclear Research Center Demokritos (NRCD), GR-15310 Attiki, Greece}
\end{center}\end{sloppypar}
\vspace{2mm}
\begin{sloppypar}
\noindent
A.~Blondel,$^{12}$
\mbox{J.-C.~Brient},
F.~Machefert,
A.~Roug\'{e},
M.~Swynghedauw,
R.~Tanaka
\linebreak
H.~Videau
\nopagebreak
\begin{center}
\parbox{15.5cm}{\sl\samepage
Laoratoire Leprince-Ringuet, Ecole
Polytechnique, IN$^{2}$P$^{3}$-CNRS, \mbox{F-91128} Palaiseau Cedex, France}
\end{center}\end{sloppypar}
\vspace{2mm}
\begin{sloppypar}
\noindent
V.~Ciulli,
E.~Focardi,
G.~Parrini
\nopagebreak
\begin{center}
\parbox{15.5cm}{\sl\samepage
Dipartimento di Fisica, Universit\`a di Firenze, INFN Sezione di Firenze,
I-50125 Firenze, Italy}
\end{center}\end{sloppypar}
\vspace{2mm}
\begin{sloppypar}
\noindent
A.~Antonelli,
M.~Antonelli,
G.~Bencivenni,
F.~Bossi,
G.~Capon,
V.~Chiarella,
P.~Laurelli,
G.~Mannocchi,$^{5}$
G.P.~Murtas,
L.~Passalacqua
\nopagebreak
\begin{center}
\parbox{15.5cm}{\sl\samepage
Laboratori Nazionali dell'INFN (LNF-INFN), I-00044 Frascati, Italy}
\end{center}\end{sloppypar}
\vspace{2mm}
\begin{sloppypar}
\noindent
J.~Kennedy,
J.G.~Lynch,
P.~Negus,
V.~O'Shea,
A.S.~Thompson
\nopagebreak
\begin{center}
\parbox{15.5cm}{\sl\samepage
Department of Physics and Astronomy, University of Glasgow, Glasgow G12
8QQ,United Kingdom$^{10}$}
\end{center}\end{sloppypar}
\vspace{2mm}
\begin{sloppypar}
\noindent
S.~Wasserbaech
\nopagebreak
\begin{center}
\parbox{15.5cm}{\sl\samepage
Utah Valley State College, Orem, UT 84058, U.S.A.}
\end{center}\end{sloppypar}
\vspace{2mm}
\begin{sloppypar}
\noindent
R.~Cavanaugh,$^{4}$
S.~Dhamotharan,$^{21}$
C.~Geweniger,
P.~Hanke,
V.~Hepp,
E.E.~Kluge,
G.~Leibenguth,
A.~Putzer,
H.~Stenzel,
K.~Tittel,
M.~Wunsch$^{19}$
\nopagebreak
\begin{center}
\parbox{15.5cm}{\sl\samepage
Kirchhoff-Institut f\"ur Physik, Universit\"at Heidelberg, D-69120
Heidelberg, Germany$^{16}$}
\end{center}\end{sloppypar}
\vspace{2mm}
\begin{sloppypar}
\noindent
R.~Beuselinck,
W.~Cameron,
G.~Davies,
P.J.~Dornan,
M.~Girone,$^{1}$
R.D.~Hill,
N.~Marinelli,
J.~Nowell,
S.A.~Rutherford,
J.K.~Sedgbeer,
J.C.~Thompson,$^{14}$
R.~White
\nopagebreak
\begin{center}
\parbox{15.5cm}{\sl\samepage
Department of Physics, Imperial College, London SW7 2BZ,
United Kingdom$^{10}$}
\end{center}\end{sloppypar}
\vspace{2mm}
\begin{sloppypar}
\noindent
V.M.~Ghete,
P.~Girtler,
E.~Kneringer,
D.~Kuhn,
G.~Rudolph
\nopagebreak
\begin{center}
\parbox{15.5cm}{\sl\samepage
Institut f\"ur Experimentalphysik, Universit\"at Innsbruck, A-6020
Innsbruck, Austria$^{18}$}
\end{center}\end{sloppypar}
\vspace{2mm}
\begin{sloppypar}
\noindent
E.~Bouhova-Thacker,
C.K.~Bowdery,
D.P.~Clarke,
G.~Ellis,
A.J.~Finch,
F.~Foster,
G.~Hughes,
R.W.L.~Jones,
M.R.~Pearson,
N.A.~Robertson,
M.~Smizanska
\nopagebreak
\begin{center}
\parbox{15.5cm}{\sl\samepage
Department of Physics, University of Lancaster, Lancaster LA1 4YB,
United Kingdom$^{10}$}
\end{center}\end{sloppypar}
\vspace{2mm}
\begin{sloppypar}
\noindent
O.~van~der~Aa,
C.~Delaere,$^{28}$
V.~Lemaitre$^{29}$
\nopagebreak
\begin{center}
\parbox{15.5cm}{\sl\samepage
Institut de Physique Nucl\'eaire, D\'epartement de Physique, Universit\'e Catholique de Louvain, 1348 Louvain-la-Neuve, Belgium}
\end{center}\end{sloppypar}
\vspace{2mm}
\begin{sloppypar}
\noindent
U.~Blumenschein,
F.~H\"olldorfer,
K.~Jakobs,
F.~Kayser,
K.~Kleinknecht,
A.-S.~M\"uller,
B.~Renk,
H.-G.~Sander,
S.~Schmeling,
H.~Wachsmuth,
C.~Zeitnitz,
T.~Ziegler
\nopagebreak
\begin{center}
\parbox{15.5cm}{\sl\samepage
Institut f\"ur Physik, Universit\"at Mainz, D-55099 Mainz, Germany$^{16}$}
\end{center}\end{sloppypar}
\vspace{2mm}
\begin{sloppypar}
\noindent
A.~Bonissent,
P.~Coyle,
C.~Curtil,
A.~Ealet,
D.~Fouchez,
P.~Payre,
A.~Tilquin
\nopagebreak
\begin{center}
\parbox{15.5cm}{\sl\samepage
Centre de Physique des Particules de Marseille, Univ M\'editerran\'ee,
IN$^{2}$P$^{3}$-CNRS, F-13288 Marseille, France}
\end{center}\end{sloppypar}
\vspace{2mm}
\begin{sloppypar}
\noindent
F.~Ragusa
\nopagebreak
\begin{center}
\parbox{15.5cm}{\sl\samepage
Dipartimento di Fisica, Universit\`a di Milano e INFN Sezione di
Milano, I-20133 Milano, Italy.}
\end{center}\end{sloppypar}
\vspace{2mm}
\begin{sloppypar}
\noindent
A.~David,
H.~Dietl,
G.~Ganis,$^{27}$
K.~H\"uttmann,
G.~L\"utjens,
W.~M\"anner,
\mbox{H.-G.~Moser},
R.~Settles,
M.~Villegas,
G.~Wolf
\nopagebreak
\begin{center}
\parbox{15.5cm}{\sl\samepage
Max-Planck-Institut f\"ur Physik, Werner-Heisenberg-Institut,
D-80805 M\"unchen, Germany\footnotemark[16]}
\end{center}\end{sloppypar}
\vspace{2mm}
\begin{sloppypar}
\noindent
J.~Boucrot,
O.~Callot,
M.~Davier,
L.~Duflot,
\mbox{J.-F.~Grivaz},
Ph.~Heusse,
A.~Jacholkowska,$^{6}$
L.~Serin,
\mbox{J.-J.~Veillet}
\nopagebreak
\begin{center}
\parbox{15.5cm}{\sl\samepage
Laboratoire de l'Acc\'el\'erateur Lin\'eaire, Universit\'e de Paris-Sud,
IN$^{2}$P$^{3}$-CNRS, F-91898 Orsay Cedex, France}
\end{center}\end{sloppypar}
\vspace{2mm}
\begin{sloppypar}
\noindent
P.~Azzurri, 
G.~Bagliesi,
T.~Boccali,
L.~Fo\`a,
A.~Giammanco,
A.~Giassi,
F.~Ligabue,
A.~Messineo,
F.~Palla,
G.~Sanguinetti,
A.~Sciab\`a,
P.~Spagnolo
R.~Tenchini
A.~Venturi
P.G.~Verdini
\samepage
\begin{center}
\parbox{15.5cm}{\sl\samepage
Dipartimento di Fisica dell'Universit\`a, INFN Sezione di Pisa,
e Scuola Normale Superiore, I-56010 Pisa, Italy}
\end{center}\end{sloppypar}
\vspace{2mm}
\begin{sloppypar}
\noindent
O.~Awunor,
G.A.~Blair,
G.~Cowan,
A.~Garcia-Bellido,
M.G.~Green,
L.T.~Jones,
T.~Medcalf,
A.~Misiejuk,
J.A.~Strong,
P.~Teixeira-Dias
\nopagebreak
\begin{center}
\parbox{15.5cm}{\sl\samepage
Department of Physics, Royal Holloway \& Bedford New College,
University of London, Egham, Surrey TW20 OEX, United Kingdom$^{10}$}
\end{center}\end{sloppypar}
\vspace{2mm}
\begin{sloppypar}
\noindent
R.W.~Clifft,
T.R.~Edgecock,
P.R.~Norton,
I.R.~Tomalin,
J.J.~Ward
\nopagebreak
\begin{center}
\parbox{15.5cm}{\sl\samepage
Particle Physics Dept., Rutherford Appleton Laboratory,
Chilton, Didcot, Oxon OX11 OQX, United Kingdom$^{10}$}
\end{center}\end{sloppypar}
\vspace{2mm}
\begin{sloppypar}
\noindent
\mbox{B.~Bloch-Devaux},
D.~Boumediene,
P.~Colas,
B.~Fabbro,
E.~Lan\c{c}on,
\mbox{M.-C.~Lemaire},
E.~Locci,
P.~Perez,
J.~Rander,
B.~Tuchming,
B.~Vallage
\nopagebreak
\begin{center}
\parbox{15.5cm}{\sl\samepage
CEA, DAPNIA/Service de Physique des Particules,
CE-Saclay, F-91191 Gif-sur-Yvette Cedex, France$^{17}$}
\end{center}\end{sloppypar}
\vspace{2mm}
\begin{sloppypar}
\noindent
N.~Konstantinidis,
A.M.~Litke,
G.~Taylor
\nopagebreak
\begin{center}
\parbox{15.5cm}{\sl\samepage
Institute for Particle Physics, University of California at
Santa Cruz, Santa Cruz, CA 95064, USA$^{22}$}
\end{center}\end{sloppypar}
\vspace{2mm}
\begin{sloppypar}
\noindent
C.N.~Booth,
S.~Cartwright,
F.~Combley,$^{25}$
P.N.~Hodgson,
M.~Lehto,
L.F.~Thompson
\nopagebreak
\begin{center}
\parbox{15.5cm}{\sl\samepage
Department of Physics, University of Sheffield, Sheffield S3 7RH,
United Kingdom$^{10}$}
\end{center}\end{sloppypar}
\vspace{2mm}
\begin{sloppypar}
\noindent
A.~B\"ohrer,
S.~Brandt,
C.~Grupen,
J.~Hess,
A.~Ngac,
G.~Prange
\nopagebreak
\begin{center}
\parbox{15.5cm}{\sl\samepage
Fachbereich Physik, Universit\"at Siegen, D-57068 Siegen, Germany$^{16}$}
\end{center}\end{sloppypar}
\vspace{2mm}
\begin{sloppypar}
\noindent
C.~Borean,
G.~Giannini
\nopagebreak
\begin{center}
\parbox{15.5cm}{\sl\samepage
Dipartimento di Fisica, Universit\`a di Trieste e INFN Sezione di Trieste,
I-34127 Trieste, Italy}
\end{center}\end{sloppypar}
\vspace{2mm}
\begin{sloppypar}
\noindent
H.~He,
J.~Putz,
J.~Rothberg
\nopagebreak
\begin{center}
\parbox{15.5cm}{\sl\samepage
Experimental Elementary Particle Physics, University of Washington, Seattle,
WA 98195 U.S.A.}
\end{center}\end{sloppypar}
\vspace{2mm}
\begin{sloppypar}
\noindent
S.R.~Armstrong,
K.~Berkelman,
K.~Cranmer,
D.P.S.~Ferguson,
Y.~Gao,$^{13}$
S.~Gonz\'{a}lez,
O.J.~Hayes,
H.~Hu,
S.~Jin,
J.~Kile,
P.A.~McNamara III,
J.~Nielsen,
Y.B.~Pan,
\mbox{J.H.~von~Wimmersperg-Toeller}, 
W.~Wiedenmann,
J.~Wu,
Sau~Lan~Wu,
X.~Wu,
G.~Zobernig
\nopagebreak
\begin{center}
\parbox{15.5cm}{\sl\samepage
Department of Physics, University of Wisconsin, Madison, WI 53706,
USA$^{11}$}
\end{center}\end{sloppypar}
\vspace{2mm}
\begin{sloppypar}
\noindent
G.~Dissertori
\nopagebreak
\begin{center}
\parbox{15.5cm}{\sl\samepage
Institute for Particle Physics, ETH H\"onggerberg, 8093 Z\"urich,
Switzerland.}
\end{center}\end{sloppypar}
}
\footnotetext[1]{Also at CERN, 1211 Geneva 23, Switzerland.}
\footnotetext[2]{Now at Fermilab, PO Box 500, MS 352, Batavia, IL 60510, USA}
\footnotetext[3]{Also at Dipartimento di Fisica di Catania and INFN Sezione di
 Catania, 95129 Catania, Italy.}
\footnotetext[4]{Now at University of Florida, Department of Physics, Gainesville, Florida 32611-8440, USA}
\footnotetext[5]{Also Istituto di Cosmo-Geofisica del C.N.R., Torino,
Italy.}
\footnotetext[6]{Also at Groupe d'Astroparticules de Montpellier, Universit\'{e} de Montpellier II, 34095, Montpellier, France.}
\footnotetext[7]{Supported by CICYT, Spain.}
\footnotetext[8]{Supported by the National Science Foundation of China.}
\footnotetext[9]{Supported by the Danish Natural Science Research Council.}
\footnotetext[10]{Supported by the UK Particle Physics and Astronomy Research
Council.}
\footnotetext[11]{Supported by the US Department of Energy, grant
DE-FG0295-ER40896.}
\footnotetext[12]{Now at Departement de Physique Corpusculaire, Universit\'e de
Gen\`eve, 1211 Gen\`eve 4, Switzerland.}
\footnotetext[13]{Also at Department of Physics, Tsinghua University, Beijing, The People's Republic of China.}
\footnotetext[14]{Supported by the Leverhulme Trust.}
\footnotetext[15]{Permanent address: Universitat de Barcelona, 08208 Barcelona,
Spain.}
\footnotetext[16]{Supported by Bundesministerium f\"ur Bildung
und Forschung, Germany.}
\footnotetext[17]{Supported by the Direction des Sciences de la
Mati\`ere, C.E.A.}
\footnotetext[18]{Supported by the Austrian Ministry for Science and Transport.}
\footnotetext[19]{Now at SAP AG, 69185 Walldorf, Germany}
\footnotetext[20]{Now at Groupe d' Astroparticules de Montpellier, Universit\'e de Montpellier II, 34095 Montpellier, France.}
\footnotetext[21]{Now at BNP Paribas, 60325 Frankfurt am Mainz, Germany}
\footnotetext[22]{Supported by the US Department of Energy,
grant DE-FG03-92ER40689.}
\footnotetext[23]{Now at Institut Inter-universitaire des hautes Energies (IIHE), CP 230, Universit\'{e} Libre de Bruxelles, 1050 Bruxelles, Belgique}
\footnotetext[24]{Also at Dipartimento di Fisica e Tecnologie Relative, Universit\`a di Palermo, Palermo, Italy.}
\footnotetext[25]{Deceased.}
\footnotetext[26]{Now at SLAC, Stanford, CA 94309, U.S.A}
\footnotetext[27]{Now at INFN Sezione di Roma II, Dipartimento di Fisica, Universit\`a di Roma Tor Vergata, 00133 Roma, Italy.}
\footnotetext[28]{Research Fellow of the Belgium FNRS}
\footnotetext[29]{Research Associate of the Belgium FNRS} 
\footnotetext[30]{Now at Liverpool University, Liverpool L69 7ZE, United Kingdom}   
\setlength{\parskip}{\saveparskip}
\setlength{\textheight}{\savetextheight}
\setlength{\topmargin}{\savetopmargin}
\setlength{\textwidth}{\savetextwidth}
\setlength{\oddsidemargin}{\saveoddsidemargin}
\setlength{\topsep}{\savetopsep}
\normalsize
\newpage
\pagestyle{plain}
\setcounter{page}{1}

\section{Introduction}

In this letter a measurement of the production
rate of \cc\ pairs from gluons in hadronic Z decays is described. The selection relies on tagging semileptonic decays of the c quarks from gluon splitting, and makes use of several discriminating variables related to the event topology.

The rate of gluon splitting to \cc\ pairs is defined as
\begin{equation}
  \gcc =\frac{N({\rm Z}\rightarrow {\rm q\bar qg}, {\rm g\rightarrow c \bar c})}{N({\rm Z}\rightarrow {\rm hadrons})} \ .
\end{equation}

Measuring \gcc\ is an important test of perturbative QCD at the Z scale. The processes \gtoc\ and \gtob\ are also significant backgrounds for several analyses involving heavy quarks.
For example, one third of the total experimental uncertainty on $R_{\rm b}$ comes from these processes~\cite{lepew}.
Furthermore, gluon splitting to heavy quarks is a background for Higgs boson searches~\cite{alephhiggs}.

The theoretical treatment of the production of heavy quarks from gluons is described in~\cite{miller,seymour,nason,likhoded} 
and the rate \gcc\ is predicted to be in the range $1.4\%$ to $2.5\%$.

Previous measurements have been performed using a 
${\rm D}^*$ tag~\cite{opal,alephD}, a lepton tag~\cite{opal,l3gcc} or event shape variables~\cite{l3gcc}. The lepton tag currently provides the most precise results.

\section{The ALEPH detector}
\label{sec:detector}

A detailed description of the ALEPH detector can be found in~\cite{detector} and of its performance in~\cite{performance}. 
A brief overview is given in this section.

Charged particles are detected in the central part of the apparatus, consisting of a high resolution silicon strip vertex detector (VDET), a cylindrical drift chamber (ITC) and a large time projection chamber (TPC). 
The three tracking detectors are immersed in a 1.5 T axial magnetic field provided by a superconducting solenoid. They are surrounded by the calorimetric system, consisting of the electromagnetic calorimeter (ECAL), the hadron calorimeter (HCAL) and the muon chambers.

The VDET~\cite{vdet} lies at the core of the tracking system. 
It is made of two layers, at average radii of 6.5 and 11.3 cm, 
each providing measurements in both the $r\phi$ and $rz$ projections, 
with a resolution of $12~\mu$m for $r\phi$ coordinates and varying between 
12 and $22~\mu$m for $z$ coordinates, depending on the track polar angle $\theta$. The angular coverage is $|{\rm cos}~\theta|<0.85$ for the inner layer and $|{\rm cos}~\theta|<0.69$ for the outer layer.

The ITC measures up to eight coordinates per track in the $r\phi$ projection, with a resolution of $150~\mu$m.

The TPC provides up to 21 three-dimensional coordinates per track, 
with resolutions in the $r\phi$ and $rz$ projections of $180~\mu$m 
and $500~\mu$m, respectively. 
The TPC also provides up to 338 measurements of the specific 
energy loss by ionization ($dE/dx$); this allows electrons to be distinguished from other charged particles by more than three standard deviations up to a momentum of 8 GeV/$c$.

The ECAL is a sampling calorimeter covering the angular range $|{\rm cos}~\theta|<0.98$, segmented in $0.9^{\circ}\times 0.9^{\circ}$ projective towers, read out in three longitudinal stacks.
The nominal thickness of the calorimeter is 22 radiation lengths.
The energy resolution for isolated electrons and photons is $\sigma_E/E = 0.009+0.18/\sqrt{E}$, with $E$ measured in GeV. 

The hadron calorimeter (HCAL) is composed of 23 layers of streamer tubes interleaved with iron slabs.
The total iron thickness corresponds to about 7 interaction lengths at normal incidence.

Electrons are identified by the characteristic longitudinal and transverse development of their associated showers in the ECAL. The $dE/dx$ information from the TPC is used to enhance the hadron rejection power, while non-prompt electrons originating from photon conversions in the detector material are rejected on the basis of their kinematical and geometrical properties.

Muons are identified by their penetration pattern in the HCAL; the additional three-dimensional coordinates measured in two double layers of external muon chambers help in resolving the remaining possible ambiguities.

The lepton identification technique is described in detail in~\cite{nancy,leptonid}, with minimum momentum cuts of 
$2$~GeV$/c$ for electrons and 
$2.5$~GeV$/c$ for muons.

\section{Preselection}
\label{sec:preselection}

The analysis is based on the LEP$\,1$ data set, which consists of about 3.9 million \zqq\ decays collected by ALEPH from 1991 to 1995.
The analysis makes also use of 8.7 million simulated \zqq\ events, 5.1 million \zbb\ events, 2.3 million \zcc\ events, and 1.8 million signal events each containing the \gtoc\ process.  
The generator is based on JETSET~7.3~\cite{jetset}, and all events are passed through a detailed simulation of the detector based on GEANT$\,3$~\cite{geant}.
Simulated events are reweighted
to take into account the latest world average of \gcc\ and \gbb.

The method relies on the analysis of events clustered into three jets, where an electron or a muon is found in the least energetic jet (taken to be the gluon jet candidate). Additional discrimination between signal and background is obtained using variables related to the event topology.

\begin{figure}[tb]
\centering
\epsfig{file=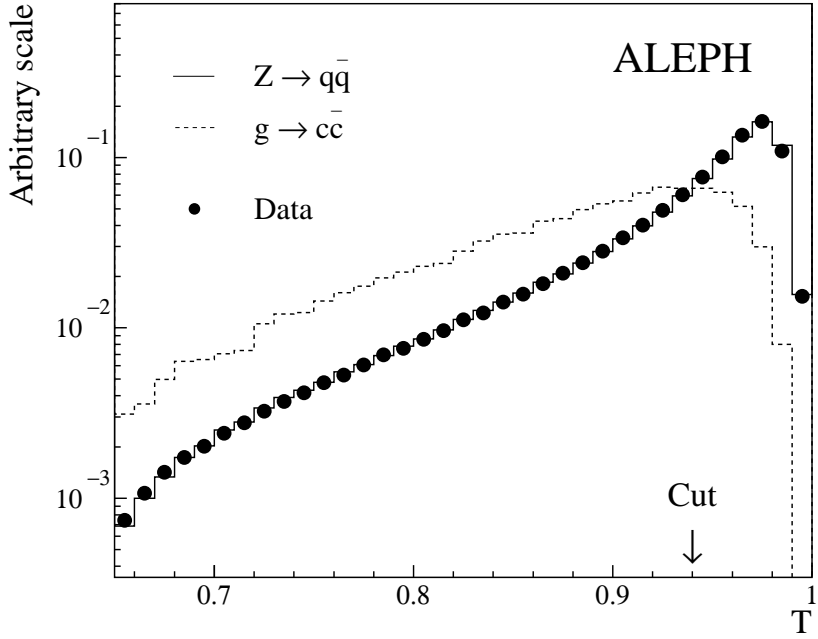, width=12 cm}
\caption{Thrust distributions for data and simulation, normalized to the same area.}
\label{fig:thrust1}
\end{figure}

First, events with a small value of the thrust are selected ($T<0.94$).
Figure~\ref{fig:thrust1} shows the thrust distribution for data and simulated hadronic events, together with the distribution of signal events.
For the events selected, particles are clustered into three jets, using the JADE algorithm~\cite{jade1} (``E scheme''). The energies of the jets are recalculated by enforcing energy and momentum conservation, under the assumptions that jet directions are perfectly measured and that jets are massless, as
\begin{equation}
\label{eq:seni}
E_i=E_{{\rm cm}}\frac{\sin\psi_{jk}}
{\sin\psi_{jk}+\sin\psi_{ij}+\sin\psi_{ik}} \ ,
\end{equation}
where $E_{{\rm cm}}$ is the centre-of-mass energy and $\psi_{ij}$ is the angle between jets $i$ and $j$.
Non-planar events, for which Eq.~\ref{eq:seni} does not hold, are rejected by requiring $\psi_{12}+\psi_{23}+\psi_{31}>358^{\circ}$.
The jet energies calculated in this way are used to order the three jets by decreasing energy.

%
Lepton candidates are searched for, following the same procedure as in~\cite{nancy}.
In the present analysis the cut on the distance of closest approach of the lepton track to the beam axis ($d_0$) is tightened to $d_0 < 0.1$~cm, to improve the rejection of leptons from K or $\pi$ decay.
Events with an identified lepton belonging to the third jet are retained for further analysis.


\begin{table}[!b] 
\centering
\caption{Composition of the preselected sample.}
\vskip 0.2cm

\begin{tabular}{|c|r|r|}
\cline{2-3}
\multicolumn{1}{c|}{$\ $} & \multicolumn{1}{c|}{e} & 
\multicolumn{1}{c|}{$\mu$}\\\hline
\gtoc \rule{0pt}{4.0mm} & $17.0\pm 0.3~\%$ & $12.8\pm 0.3~\%$\\\hline
\gtob \rule{0pt}{4.0mm} & $3.4\pm 0.2~\%$ & $2.5\pm 0.1~\%$\\\hline
\zuds \rule{0pt}{4.0mm} & $10.9\pm 0.3~\%$ & $27.7\pm 0.4~\%$\\
\zcc & $21.5\pm 0.4~\%$ & $20.0\pm 0.3~\%$\\
\zbb & $47.2\pm 0.4~\%$ & $37.0\pm 0.4~\%$\\
\hline
\end{tabular}
\label{tab:presel}
\end{table}

This preselection yields 13363 events, out of which 5639 contain an electron candidate and 7724 a muon candidate.
The composition of the two subsamples evaluated with simulated events is shown in Table~\ref{tab:presel}: the flavour content of the background is substantially different in the two cases.
The larger fraction of light quark events for the muon subsample is related to the contamination of pions. 
The fraction of different particle types contributing to the samples of muon and electron candidates is shown in Table~\ref{tab:comp_mu_el}, where prompt lepton indicates a lepton originating from the decay of a heavy flavour particle.


 \begin{table}[!ht]
 \centering
 \caption{Fraction of different particle types in the samples of muon and electron candidates, after preselection.}
\vskip 0.2cm
 \begin{tabular}{|l|r|c|l|r|}
\cline{1-2}
   \multicolumn{2}{|c|}{Muon candidates} & \multicolumn{2}{c}{$\  $}  \\
\cline{1-2}\cline{4-5}
   prompt $\mu$           & 56\% && \multicolumn{2}{|c|}{Electron candidates}  \\
\cline{1-2}\cline{4-5}
   $\pi\rightarrow \mu\nu$& 25\% &$\ \ \ \ \ \ \ \ \ \ \ \ \ $&prompt e              & 69\%  \\
\cline{1-2}\cline{4-5}
   misidentified $\pi$    & 10\% && $\gamma \rightarrow {\rm e^+e^-}$ & 26\%  \\
\cline{1-2}\cline{4-5}
   ${\rm K} \rightarrow \mu\nu$ & 5\%  && misidentified $\pi$  & 4\%   \\
\cline{1-2}\cline{4-5}
   misidentified K        & 3\%  && others                  & 1\%    \\
\cline{1-2}\cline{4-5}
   others                 & 1\%  & \multicolumn{2}{c}{$\  $}  \\
\cline{1-2}
\end{tabular}
 \label{tab:comp_mu_el}
 \end{table}


\boldmath
\section{The extraction of \gcc}
\unboldmath
\label{sec:nn}

The purity of the selected sample is inadequate to perform a measurement 
of \gcc\ with a meaningful precision. Additional variables are needed 
to provide further separation between the signal and the $\Z \to \qqbar$ 
background, as follows.

\begin{itemize}

\item After the preselection cut mentioned in Section~\ref{sec:preselection}, as 
clearly visible from the distributions in Fig.~\ref{fig:thrust1}, the thrust $T$ retains 
some discriminating power between the (multi-jet-like) signal events and 
the bulk of the $\Z \to \qqbar$ background.

\item Each event is divided into two hemispheres by a plane perpendicular
to the thrust axis. The confidence levels $B_1$ and $B_2$ (hereafter called
b-tag probabilities~\cite{btag}) that the charged particle tracks of each hemisphere
originate from the primary vertex provide a tag against $\Z \to \bbbar$ 
events and, to a lesser extent, $\Z \to \ccbar$ events.

\item The b-tag probability $b_\ell$ and the momentum $P_\ell$ of the tagged 
lepton in the third jet also contribute to the aforementioned anti-b-tag 
capability.

\item Conversely, the projection $P^3_{\rm miss}$ of the missing momentum
along the axis of the third jet is largest for ${\rm c} \to \ell$ decays, 
and is therefore discriminant against $\Z \to \uubar, \ddbar$ and \ssbar\ 
events, for which the missing momentum direction is mostly random. This 
variable is statistically almost as powerful as the third jet mass, used for 
instance in~\cite{opal,l3gcc}. However, the latter is not included in the present 
analysis because it is found to depend on details of the fragmentation 
model and to be inadequately reproduced by the simulation.

\item Finally, the relative discriminating power of the variables
mentioned above varies with the third jet polar angle. The cosine 
$\vert \cos \theta_\ell \vert$ of the tagged lepton polar angle is
therefore added to better control this dependence.

\end{itemize}

These variables are combined with an artificial neural network into single
discriminants, \Ne\ and \Nmu, for the electron and the muon 
samples separately. The neural network is trained with half the simulated
$\Z \to \qqbar$ sample and half the simulated signal sample. The remaining
statistics, together with the $\Z \to \ccbar$ and the $\Z \to \bbbar$ 
simulated samples, are used to determine the selection efficiencies.

\begin{figure}[tb]
\centering
\epsfig{file=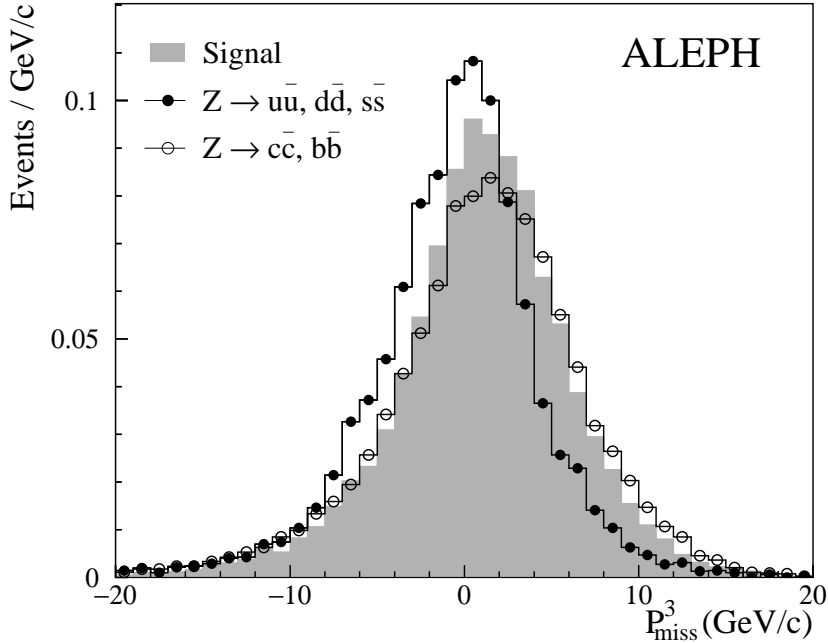, width=12cm}
\caption{Distributions of the signal and of the background components, normalized to the same area, for $P^3_{\rm miss}$.}
\label{fig:comp_pm3}
\end{figure}

\begin{figure}[!ht]
\centering
\epsfig{file=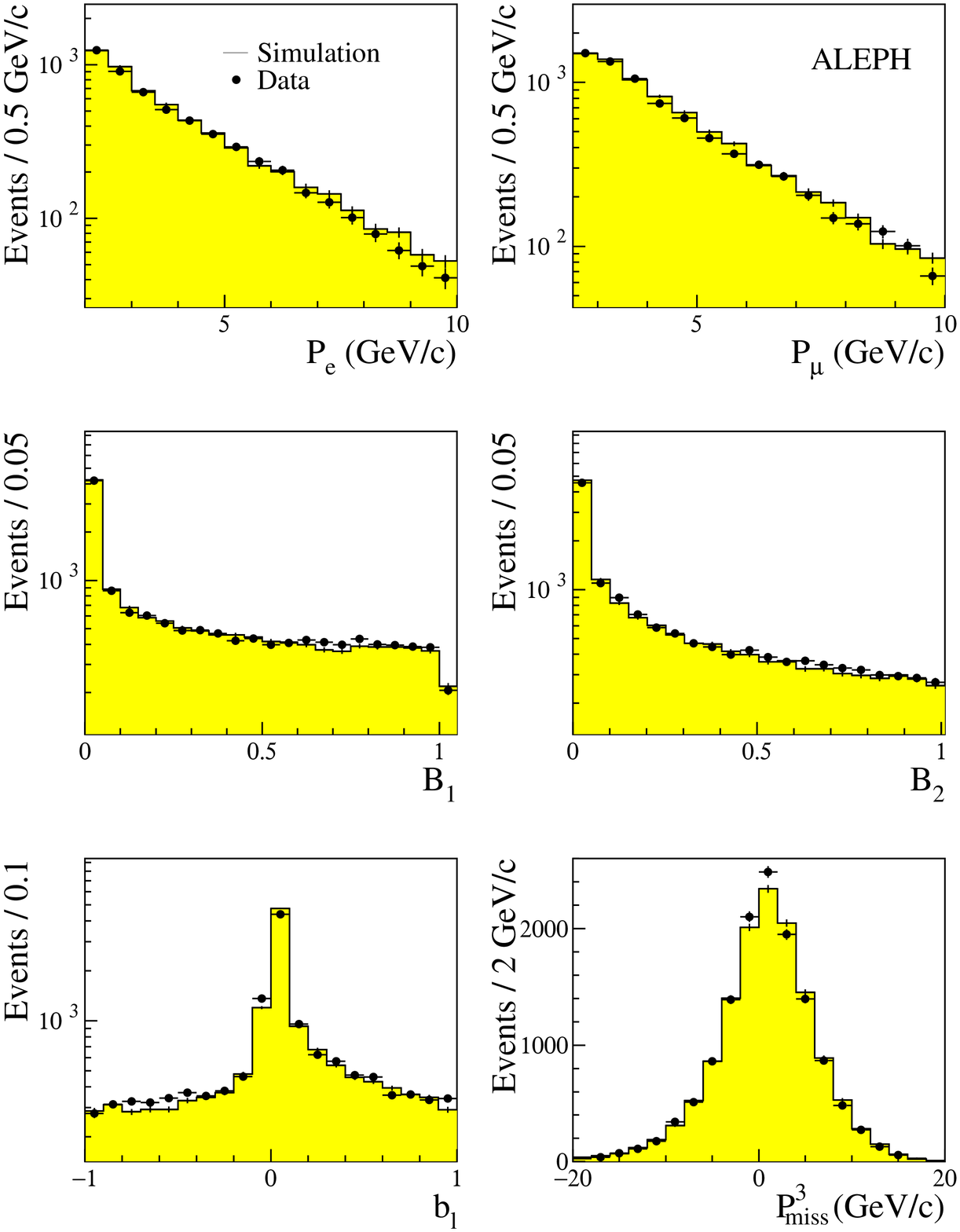, width=16.5truecm}
\caption{Distribution of the variables with highest discriminating power, in data and simulation.}
\label{fig:dtmc_tutti}
\end{figure}

As an example, the discriminant power of $P^3_{\rm miss}$ is shown in Fig.~\ref{fig:comp_pm3}, where the distributions of the signal and background components are compared (normalized to the same area).
Figure~\ref{fig:dtmc_tutti} shows the agreement between data and simulation for the variables with highest discriminating power, at preselection level.

\begin{figure}[tb]
\centering
\epsfig{file=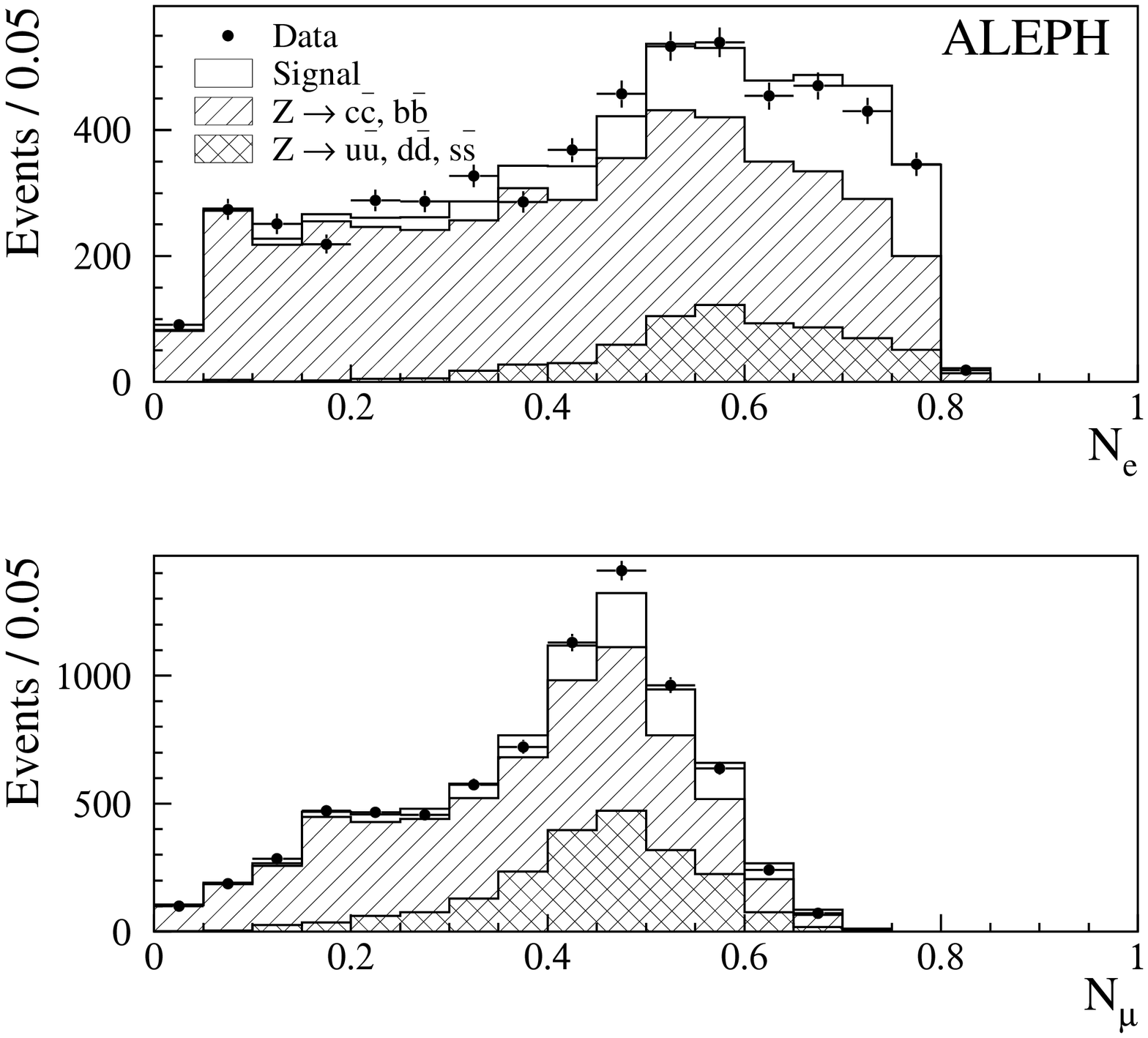, width=12.5cm}
\caption{Output of the neural network for electrons and muons. 
In these plots, the signal is normalized according to the results 
obtained from Eq.~\ref{eq:gcc}.}
\label{fig:output}
\end{figure}

The distributions of \Ne\ and \Nmu\ are shown in Fig.~\ref{fig:output}.
The separation between signal and background is worse for the muon subsample, due to the high contamination of non-prompt muons in the preselected sample.

The cuts on \Ne\ and \Nmu\ are chosen to be 0.55 and 0.45 in order to 
minimize the total uncertainty on \gcc.
This final selection cut yields 2258 events with an electron candidate and 
3332 events with a muon candidate. 
The sample composition is estimated from the simulation, 
and is shown in Table~\ref{tab:sel}.

The value of the gluon splitting rate is extracted as
\begin{equation}
\gcc = \frac{f \, - \,(1 - \gbb ) \, \epsilon_Q \, -  \gbb \, \epsilon_B}
{\epsilon_C \, - \, \epsilon_Q} \ ,
\label{eq:gcc}
\end{equation}
where $f$ is the fraction of events selected in the data. 
The selection efficiencies $\epsilon_Q$, $\epsilon_C$, $\epsilon_B$ 
for events with, respectively, no gluon splitting, gluon splitting in \ccbar, 
and gluon splitting in \bbbar\ are estimated from the simulation 
(Table~\ref{tab:seleff}).

\begin{table}[!h] 
\centering
\caption{Composition of the selected sample, after the cuts on \Ne\ and \Nmu.}
\vskip 0.2cm
\begin{tabular}{|c|r|r|}
\cline{2-3}
\multicolumn{1}{c|}{$\ $} & \multicolumn{1}{c|}{e} & 
\multicolumn{1}{c|}{$\mu$}\\\hline
\gtoc \rule{0pt}{4.0mm} & $25~\%$ & $18~\%$\\\hline
\gtob \rule{0pt}{4.0mm} & $4~\%$ & $2~\%$\\\hline
\zuds \rule{0pt}{4.0mm} & $18~\%$ & $35~\%$\\
\zcc & $28~\%$ & $22~\%$\\
\zbb & $25~\%$ & $23~\%$\\
\hline
\end{tabular}
\label{tab:sel}
\end{table}

\begin{table}[!b] 
\centering
\caption{Fraction of events selected and selection efficiencies for the different event categories, with statistical errors.}
\vskip 0.2cm
\begin{tabular}{|c|c|c|}
\cline{2-3}
\multicolumn{1}{c|}{$\ $}& electrons $(\%)$ & muons $(\%)$\\\hline
$f$ & $0.0600\pm 0.0013$ & $0.0885\pm 0.0015$ \\\hline
$\epsilon_Q$ & $0.0431\pm 0.0006$ & $0.0747\pm 0.0009$\\
$\epsilon_C$ & $0.491\pm 0.005$ & $0.473\pm 0.005$\\
$\epsilon_B$ & $0.81\pm 0.06$ & $0.80\pm 0.06$\\\hline
\end{tabular}

\label{tab:seleff}
\end{table}

In the calculation of \gcc, \gbb\ is fixed to the latest world 
average value \mbox{$(2.54\pm 0.51)\times 10^{-3}$}~\cite{lephf}. 
The results extracted from the two subsamples are 
\mbox{$\gcc^{\rm e}=(3.32\pm 0.28)\%$} and 
\mbox{$\gcc^{\mu}=(2.99\pm 0.38)\%$}, 
where the errors only account for the statistical uncertainty on $f$.

\section{Systematic errors}
\label{sec:syst}

The sources of uncertainty on the selection efficiencies given in Table~\ref{tab:seleff} are discussed in this section.
The resulting systematic uncertainties on \gcc\ are summarized in Table~\ref{tab:syst}.
\renewcommand{\theenumi}{\roman{enumi}}
\begin{enumerate}

\item {\em Statistics of the simulation} \newline
The statistical uncertainties on the selection efficiencies 
translate to \mbox{$\Delta \gcc^{\rm e}=\pm 0.13\%$} and 
\mbox{$\Delta \gcc^{\mu}=\pm 0.21\%$}.

\item {\em Gluon splitting to $b\bar b$ pairs} \newline
The uncertainty on \gbb\ yields $\Delta \gcc^{\rm e}=\pm 0.09\%$ 
and $\Delta \gcc^{\mu}=\pm 0.10\%$.

\item {\em Thrust cut} \newline
The efficiency of the thrust cut is found to be slightly but significantly different in data and simulation.
The effect is taken into account by reweighting the simulated efficiencies to the data efficiencies. If no reweighting is performed, a shift is observed in the result: 
$\Delta \gcc^{\rm e}=+ 0.07\%$ and 
$\Delta \gcc^{\mu}=+ 0.11\%$. 
This shift is taken as an error estimate, with full correlation between the two channels.

\item {\em Heavy quark properties} \newline
The lifetimes of the b hadrons determine the impact parameter 
distribution of the decay products,
which is the basis of the ``anti-b tag'' method used in this analysis. Their experimental values are taken from~\cite{pdg}; the uncertainties translate to 
$\Delta \gcc^{\rm e}=\pm 0.03\%$ and $\Delta \gcc^{\mu}=\pm 0.03\%$.

The semileptonic branching ratios of c and b hadrons~\cite{alessia} 
affect the flavour composition of the samples. 
Their uncertainties and their influence on the result of the analysis 
are shown in Table~\ref{tab:br_spectra}.

\begin{table}[!b]
\centering
\caption{Effect of the uncertainties on the rates and energy spectra of heavy hadron semileptonic decays.}
\vskip 0.2cm
\begin{tabular}{|l|c|c|}
\cline{2-3}
\multicolumn{1}{c|}{$\ $} & $\Delta \gcc^{\rm e}(\%)$ & $\Delta \gcc^{\mu}(\%)$\\\hline
${\rm BR}(\bl)=(10.65\pm 0.23)\%$ & $\pm 0.05$ & $\pm 0.06$\\
${\rm BR}(\bcl)=(8.04\pm 0.19)\%$ & $\pm 0.02$ & $\pm 0.03$\\
${\rm BR}(\cl)=(9.73\pm 0.32)\%$ & $\pm 0.23$ & $\pm 0.20$\\
\hline
$\bl$ modelling & $-0.01$ & $< 0.01$\\
$\bcl$ modelling & $+0.02$ & $+0.02$\\
$\cl$ modelling & $+0.03$ & $+0.12$\\
\hline
\end{tabular}

\label{tab:br_spectra}
\end{table}

The energy spectra of leptons coming from 
$\bl$, $\cl$ and $\bcl$ decays have been tuned and varied as in~\cite{asym}, and the effect has been propagated to the measured gluon splitting rate as shown in Table~\ref{tab:br_spectra}.

The $x_E~(\equiv p_{{\rm had}}/E_{{\rm beam}})$ distribution of the heavy hadrons affects their decay length distribution and the momenta of their decay products.
The uncertainty coming from the limited knowledge of this quantity has been evaluated by varying the mean \xEb\ and \xEc\ 
within their estimated errors: \mbox{$\langle \xEb \rangle = 0.702 \pm 0.008$},
\mbox{$\langle \xEc \rangle = 0.484 \pm 0.008$}~\cite{lephf}, 
obtaining \mbox{$\Delta \gcc^{\rm e}=\pm 0.02\%$}, 
\mbox{$\Delta \gcc^{\mu}=\pm 0.03\%$} and 
\mbox{$\Delta \gcc^{\rm e}=\pm 0.04\%$}, 
\mbox{$\Delta \gcc^{\mu}=\pm 0.05\%$}, respectively.

\item {\em Lepton identification efficiency} \newline
  The charm particles produced from gluon splitting are generally of low energy
and this has two consequences for the lepton selection. Close to the lower
momentum cut-offs the identification efficiency for  the electrons degrades due
to the poorer relative energy resolution of the calorimeter, whilst for the 
muons
there is an increasing level of backgrounds from $\pi$ and K decays, as discussed in (vi).
The systematic error associated to the lepton identification efficiency 
is estimated as in~\cite{nancy}. The resulting effect is 
$\Delta \gcc^{\rm e}=\pm 0.18\%$ and 
$\Delta \gcc^{\mu}=\pm 0.01\%$.

\item {\em Fake and non-prompt leptons} \newline
A light hadron (K, $\pi$), or a light hadron decaying to a muon 
within the tracking volume (${\rm K}\rightarrow\mu$, $\pi\rightarrow\mu$), 
can be selected by the muon identification algorithm with a certain 
probability. A reliable estimate of such a mistag probability is 
important for the evaluation of the purity of the selected sample, 
derived from simulated events.

Two high-purity samples of light hadrons are selected, 
and the muon selection efficiencies in data and simulation are compared.

The first sample is selected inclusively by means of a dedicated 
``uds'' tag, designed to select event hemispheres that do not contain
decay products of heavy flavoured particles. 
Such a tag is based on the presence 
of secondary vertices, on the momentum of the fastest
charged particle, on the total visible energy and, if a lepton candidate 
is present, on its transverse momentum with respect to the jet axis. 
Tracks are selected 
in the opposite hemisphere with the muon identification kinematic
cuts as  in 
Section~\ref{sec:preselection}.

A second sample is obtained by reconstructing ${\rm K}^0_{\rm S}$ decays, 
as in~\cite{k0s}. The candidates are required to have a decay length 
larger than 1~cm, and an invariant mass in a window of width 
$(0.014+0.3 P_{\rm K}/\sqrt{s})$~GeV$/c^2$ around the nominal ${\rm K}^0_{\rm S}$ mass. 
The contamination of prompt muons from heavy flavour decays is further reduced 
by applying a soft uds tag to the hemisphere opposite to the ${\rm K}^0_{\rm S}$
candidate.

The number of tracks selected and identified as muon in data and simulation are shown, for the two samples, in Tables~\ref{tab:fake1} and~\ref{tab:fake2}.
The mistag probability in the data is calculated after subtracting the prompt 
muon component estimated from the simulation.

\begin{table}[!t] 
\centering
\caption{Performance of the prompt-muon identification on uds-fragmentation hadrons.}
\vskip 0.2cm
\begin{tabular}{|c|c|c|}
\cline{2-3}
\multicolumn{1}{c|}{$\ $}  & Simulation & Data\\
\hline
Number of tracks & 564396 & 332103\\
Number of prompt $\mu$ & 432 & --- \\
\hline
Tracks identified as $\mu$ & 3072 & 2077 \\
Prompt $\mu$ identified as $\mu$ & 351 & --- \\
\hline
Mistag $(10^{-3})$ \rule{0pt}{4.0mm} & $4.83\pm 0.08$ & $5.64\pm 0.13$ \\
\hline
\end{tabular}
\label{tab:fake1}
\end{table}

\begin{table}[!ht] 
\centering
\caption{Performance of the prompt-muon identification on pions from 
${\rm K}^0_{\rm S}$ decays.}
\vskip 0.2cm
\begin{tabular}{|c|c|c|}
\cline{2-3}
\multicolumn{1}{c|}{$\ $} & Simulation & Data\\
\hline
Number of tracks & 113068 & 50162\\
Number of prompt $\mu$ & 199 & --- \\
\hline
Tracks identified as $\mu$ & 662 & 326 \\
Prompt $\mu$ identified as $\mu$ & 156 & --- \\
\hline
Mistag $(10^{-3})$ \rule{0pt}{4.0mm} & $4.48\pm 0.20$ & $5.13\pm 0.31$ \\
\hline
\end{tabular}
\label{tab:fake2}
\end{table}

The two samples consistently indicate that the mistag probability is higher in data than in the simulation, by the ratios $1.17\pm 0.03$ and $1.15\pm 0.09$. 
A correction factor of $1.17\pm 0.05$ has been conservatively applied to the simulation.
The resulting uncertainty on the measurement is $\Delta \gcc^{\mu}=\pm 0.50\%$.

For electrons an uncertainty of $20\%$ is assigned to the rate of misidentified hadrons, and of $5\%$ to the rate of photon conversions~\cite{nancy}, yielding \mbox{$\Delta \gcc^{\rm e}=\pm 0.09\%$}.

\item{\em Generators} \newline
The modelling of the \gtoc\ process affects the selection efficiencies and hence the value of \gcc.
The process can be described in terms of three basic variables: the energy 
$E_{\rm g}$ of the gluon, its effective mass $m_{\rm g}^*$ 
and the decay angle $\theta^*$ of the c quark, 
measured in the gluon rest frame.
The distributions of these variables as given by the JETSET generator are reweighted to match the prediction of the HERWIG generator~\cite{herwig1}, and the difference observed in the measured value of \gcc\ is taken as an estimate of the systematic uncertainty. The procedure yields 
$\Delta \gcc^{\rm e}=+ 0.05\%$ and 
$\Delta \gcc^{\mu}=+ 0.26\%$ for  the effect of $E_{\rm g}$ and $m_{\rm g}^*$ 
(which are strongly correlated),
$\Delta \gcc^{\rm e}=- 0.06\%$ and 
$\Delta \gcc^{\mu}=- 0.12\%$ for the angular distribution.

\item{\em Mass of the charm quark} \newline
The mass of the charm quark is taken to be $1.2 \pm 0.2$ GeV/$c^2$. 
A shift in the charm mass results in a variation of the $E_{\rm g}$ and 
$m_{\rm g}^*$ distributions. The corresponding uncertainties 
on the measured values are $\Delta \gcc^{\rm e}=\pm 0.19\%$ and 
$\Delta \gcc^{\mu}=\pm 0.24\%$.

\end{enumerate}

\begin{table}[!ht]
\centering
\caption{Summary of the statistical and systematic uncertainties.}
\vskip 0.2cm
\begin{tabular}{|l|c|c|}
\hline
{\bf Source of error} \rule{0pt}{4.0mm} & $\Delta \gcc^{\rm e} \; (\%)$ 
& $\Delta \gcc^{\mu} \; (\%)$\\[0.1cm]\hline\hline
{\bf Statistical error} & $ 0.28$ & $ 0.38$\\\hline
Statistics of the simulation & $ 0.13$ & $ 0.21$\\
\gbb & $ 0.09$ & $ 0.10$\\
Thrust cut efficiency & 0.07 & 0.11 \\
BR($\bl$) & $ 0.05$ & $ 0.06$\\
BR($\bcl$) & $ 0.02$ & $ 0.03$\\
BR($\cl$) & $ 0.23$ & $ 0.20$\\
$\bl$ spectrum & 0.01 & $< 0.01$\\
$\bcl$ spectrum & 0.02 & 0.02\\
$\cl$ spectrum & 0.03 & 0.12\\
Lifetimes of b hadrons & $ 0.03$ & $ 0.03$\\
$\langle \xEc \rangle$ & $0.04$ & $0.05$\\
$\langle \xEb \rangle$ & $0.02$ & $0.03$\\
Lepton identification & $ 0.18$ & $ 0.01$\\
Lepton background & $ 0.09$ & $ 0.50$\\
Generators, $\theta^{*}$ & 0.06 & 0.12 \\
Generators, $E_{\rm g},m_{\rm g}^*$ & 0.05 & 0.26 \\
$m_{\rm c}$ & 0.19 & 0.24 \\
\hline
{\bf Total systematic error} & $ 0.42$ & $ 0.72$\\\hline
\end{tabular}
\label{tab:syst}
\end{table}

\section{Consistency checks}
\label{sec:checks}

\subsection{Stability of the neural network cuts}

\begin{figure}[tb]
\centering
\epsfig{file=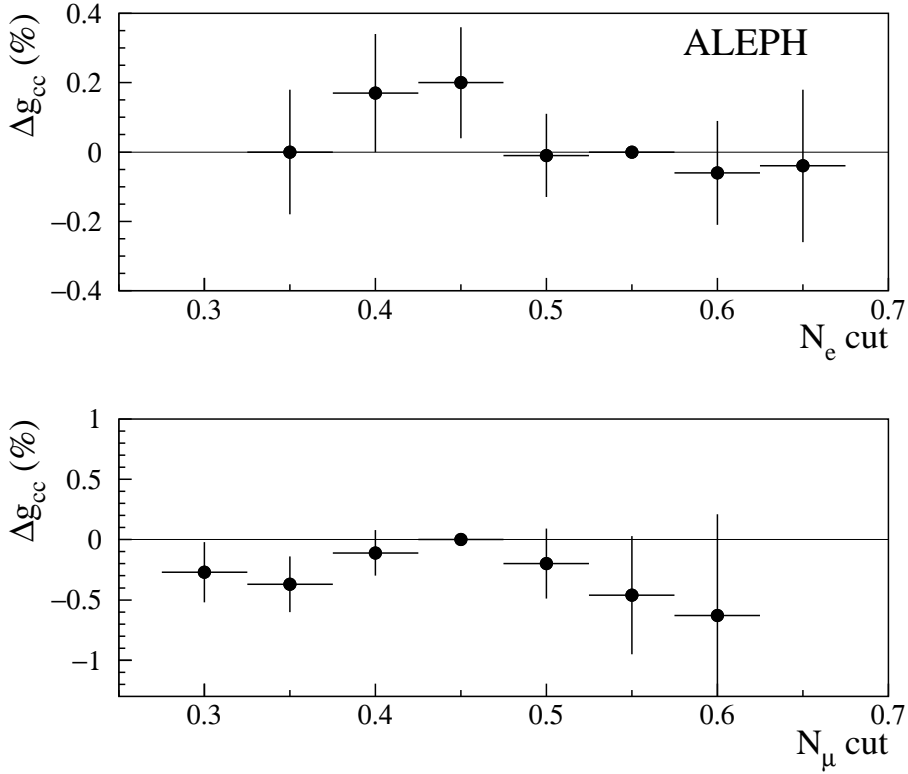, width=12.5cm }
\caption{$\Delta \gcc$ versus the \Ne, \Nmu\ cut.
The error bars indicate the uncorrelated uncertainties.}
\label{fig:stabnn}
\end{figure}

The stability of the results is checked against the cuts on \Ne\ and \Nmu.
Figure~\ref{fig:stabnn} shows, for the electron and muon subsamples, the variation of the results versus the cuts, together with the statistically uncorrelated error. No significant dependence is observed.

\subsection{Analysis of simulated events}

In the simulation, without reweighting, the fraction of hadronic events with a gluon splitting to a
\ccbar\ pair is $1.79\%$. 
The analysis applied to simulated events yields 
$\gcc^{\rm e}=(1.70\pm 0.21)\%$ and 
$\gcc^{\mu}=(1.75\pm 0.29)\%$, consistent with the input value.

\boldmath
\subsection{Shape of \Ne\ and \Nmu}
\unboldmath

The shape of the neural network output distributions for the electron
and the muon samples is shown in Fig.~\ref{fig:output}. The agreement between data 
and simulation over the whole range confirms that the excess observed in 
the data is compatible with originating from gluon splitting to \ccbar. 
A fit of \gcc\ to this shape would have therefore been feasible and would 
have (marginally) improved the statistical accuracy of the measurement. 
However, because this shape is expected to be sensitive to details 
of the fragmentation simulation, the already dominant systematic 
uncertainty would have increased accordingly, thus reducing the 
significance of the measurement.

\section{Results and conclusions}
\label{sec:conclusions}

The rate of gluon splitting to a \ccbar\ pair in hadronic Z decays has been measured from samples containing an electron or a muon candidate:
\begin{eqnarray}
\gcc^{\rm e}&=&\left(3.32\pm 0.28 \rm{(stat)}\pm 0.42 \rm{(syst)}\right)\%\;,\nonumber \\
\gcc^{\mu}&=&\left(2.99\pm 0.38 \rm{(stat)}\pm 0.72 \rm{(syst)}\right)\%\;.\nonumber
\end{eqnarray}

The two results are combined with the correlations between systematic uncertainties taken into account using the BLUE technique~\cite{blue}. The combined final ALEPH result is
\begin{eqnarray}
\gcc&=&\left(3.26\pm 0.23 \rm{(stat)}\pm 0.42 \rm{(syst)}\right)\%.\nonumber
\end{eqnarray}

This result is in agreement with the other measurements~\cite{opal,alephD,l3gcc} and with the present world average $\gcc=(2.96\pm 0.38)\%$~\cite{lephf}.



\section*{Acknowledgements}

We wish to thank our colleagues from the accelerator divisions for the successful operation of LEP. It is also a pleasure to thank the engineers and technicians of the collaborating institutions for their support in constructing and maintaining the ALEPH experiment. Those of us from non-member states thank CERN for its hospitality.


\begin{thebibliography}{99}
\bibitem{lepew} The LEP Electroweak Working Group, \emph{A combination of preliminary electroweak measurements and constraints on the Standard Model} (2002) CERN-EP/2002-091.
\bibitem{alephhiggs} The ALEPH collaboration, \emph{Final results of the searches for neutral Higgs bosons in $e^+e^-$ collisions at $\sqrt{s}$ up to 209 GeV}, Phys. Lett. {\bf B526} (2002) 191.
\bibitem{miller} D.J. Miller, \emph{The multiplicity of heavy quark pairs from gluon splitting in $e^+ e^-$ annihilation}, 1998, RAL-TR-98-043, hep-ph/9806302.
\bibitem{seymour} M.H. Seymour, \emph{Heavy quark pair multiplicity in  $e^+ e^-$ events}, Nucl. Phys. {\bf B436}, (1995) 163.
\bibitem{nason} S. Frixione et al., \emph{Heavy quark production}, Adv. Ser. Direct. High Energy Physics {\bf 15} (1998) 609.
\bibitem{likhoded} A.K. Likhoded, A.I. Onishchenko, \emph{Gluon splitting into a pair of heavy quarks in $e^+e^-$ annihilation}, Phys. Atom. Nucl. {\bf 60} (1997) 623. 
\bibitem{opal} The OPAL collaboration, \emph{Measurement of the production rate of charm quark pairs from gluons in hadronic Z decays}, Eur. Phys. J. {\bf C13} (2000) 1.
\bibitem{alephD} The ALEPH collaboration, \emph{Study of charm production in Z decays}, Eur. Phys. J. {\bf C16} (2000) 597.
\bibitem{l3gcc} The L3 collaboration, \emph{Measurement of the probability of gluon splitting into charmed quarks in hadronic Z decays},  Phys. Lett. {\bf B476} (2000) 243.
\bibitem{detector} The ALEPH collaboration, \emph{ALEPH: a detector for electron-positron annihilations at LEP}, Nucl. Instrum. and Methods {\bf A294} (1990) 121.
\bibitem{performance} The ALEPH collaboration, \emph{Performance of the ALEPH detector at LEP}, Nucl. Instrum. and Methods {\bf A360} (1995) 481.
\bibitem{vdet} B. Mours et al., \emph{The design, construction and performance of the ALEPH silicon vertex detector}, Nucl. Instrum. and Methods {\bf A379} (1996) 101.
\bibitem{nancy} The ALEPH collaboration, \emph{Inclusive semileptonic branching ratios of b hadrons produced in Z decays}, Eur. Phys. J. {\bf C22} (2002) 613.
\bibitem{leptonid} The ALEPH collaboration, \emph{Heavy quark tagging with leptons in the ALEPH detector}, Nucl. Instrum. and Methods {\bf A346} (1994) 461.
\bibitem{jetset} T. Sj\"ostrand, \emph{PYTHIA 5.7 and JETSET 7.4 physics and manual}, 1993, CERN-TH.7112/93.
\bibitem{geant} R. Brun et al., \emph{GEANT}, 1984, CERN DD/EE/84-1.
\bibitem{jade1} The JADE collaboration, W. Bartel et al., \emph{Experimental studies on multi-jet production in $e^+e^-$ annihilation at PETRA energies}, Z. Phys. {\bf C33} (1986) 23.
\bibitem{btag} The ALEPH collaboration, \emph{A measurement of $R_b$ using mutually exclusive tags}, Phys. Lett. {\bf B401} (1997) 163.
\bibitem{lephf} The LEP/SLD Heavy Flavour Working Group, \emph{Final input parameters for the LEP/SLD heavy flavour analyses}, 2001, LEPHF/2001-01.
\bibitem{pdg} Particle Data Group, K. Hagiwara et al., \emph{Review of Particle Physics}, Phys. Rev. {\bf D66} (2002) 1.
\bibitem{alessia} A. Tricomi, \emph{The $B$ semileptonic branching fractions in $Z\to b\bar b$ decays}, Proceedings of the 31st International Conference on High Energy Physics, Amsterdam, 24-31 July 2002.
\bibitem{asym} The ALEPH collaboration, \emph{Measurement of the forward--backward asymmetry in $Z\rightarrow b\bar b$ and $Z\rightarrow c\bar c$ decays with leptons}, Eur. Phys. J. {\bf C24} (2002) 177.
\bibitem{k0s} The ALEPH collaboration, \emph{Production of $K^0$ and $\Lambda$ in hadronic Z decays}, Z. Phys. {\bf C64} (1994) 361.
\bibitem{herwig1} G. Marchesini et al., \emph{HERWIG -- A Monte Carlo event generator for simulating Hadron Emission Reactions With Interfering Gluons}, Comp. Phys. Commun. {\bf 67} (1992) 465;\newline
G. Corcella et al., \emph{HERWIG 6.3 release note}, JHEP {\bf 0101} (2001) 010.
\bibitem{blue} L. Lyons, D. Gibaut, P. Clifford, \emph{How to combine correlated estimates of a single physical quantity}, Nucl. Instrum. and Methods {\bf A270} (1988) 110.
\end{thebibliography}
\end{document}